\documentclass[amsmath]{aastex702}
\usepackage{hyperref}

\begin{document}
\title{Disk-averaged spectrum generation for spherically symmetric planets and moons}
\correspondingauthor{N. A. Teanby}
\email{n.teanby@bristol.ac.uk}

\author[0000-0003-3108-5775]{N. A. Teanby}
\affiliation{School of Earth Sciences, University of Bristol, Wills Memorial Building, Queens Road, Bristol, BS8 1RJ, UK.}
\email{n.teanby@bristol.ac.uk}

\begin{abstract}

Here I present a simple and efficient method for generating a disk-averaged planetary spectrum. The technique uses a weighted average of component spectra generated on an irregular grid of offsets from the planet center. This accounts for variations in center to limb brightness and emission from beyond the limb, which is important for bodies with extended atmospheres such as Saturn's moon Titan. The method assumes spherically symmetric emission and requires a grid of component spectra fine enough to faithfully capture radial radiance variations. An example application to Titan's 7.7~$\mu$m methane emission demonstrates the method.

\end{abstract}


\section{Introduction} 

Observations of planets and moons in our solar system and beyond are often not spatially resolved.
This is especially common for Earth-orbiting space telescopes and ground-based facilities at infra-red and sub-mm wavelengths such as JWST, Herschel, ISO, and IRTF.
In such cases a disk-averaged spectrum must be analyzed.
If atmospheric or surface properties are being inverted from these spectra, it is critical that the radiative transfer forward model accurately accounts for the observation's disk-averaged nature.
For a general planet this requires knowledge of spatial variation of atmospheric and surface properties \citep{05tinetal,09heaetal} and for rapidly rotating planets may involve accounting for Doppler shift variation with latitude \citep{22teaetal}.
However, for a spherically symmetric planet considerable simplifications can be made, which is what I consider here.

\section{Method}

Here I follow an improved version of the approach in \citet{13teaetal}, which splits emission into concentric annuli. 
Consider a planet that emits radiance $s(x,\psi)$, where $x$ is offset distance from the planet center and $\psi$ is azimuth from north.
The disc-averaged radiance $\bar{s}$ is:
\begin{eqnarray}
\bar{s} &=& \frac{\int_0^r \int_0^{2\pi} x s(x,\psi)\,d\psi\,dx}{\pi r^2}
\end{eqnarray}
where $r$ is the distance from planet center where radiance becomes negligible (usually some distance off the planet limb), $x\,d\psi\,dx$ is the area element, and $\pi r^2$ is the total area being averaged over.
If the planet is assumed to be spherically symmetrical, there is no $\psi$ dependence and this simplifies to:
\begin{eqnarray}
\bar{s} &=& \frac{\int_0^r 2 \pi x s(x)\,dx}{\pi r^2}
\end{eqnarray}
Descretizing the integral using a grid of distances $x_i, i=0 \ldots n-1$ from the planet center:
\begin{eqnarray}
\bar{s} &=& \frac{1}{\pi r^2} \sum_{i=0}^{n-2} \int_{x_i}^{x_{i+1}} 2 \pi x s(x)\,dx
\end{eqnarray}
I now assume that radiance $s(x_i)$ varies linearly with $x$ within each interval $x_i \rightarrow x_{i+1}$ which gives the approximation:
\begin{eqnarray}
\label{eqn:approx}
\bar{s} &\approx&  \frac{1}{\pi r^2} \sum_{i=0}^{n-2} \int_{x_i}^{x_{i+1}} 2 \pi x \left[ s_i + \left(\frac{s_{i+1}-s_i}{x_{i+1}-x_i}\right)(x-x_i)\right]\,dx
\end{eqnarray}
where the term in square brackets is the standard straight line equation through two points $(x_i,s_i)$ and $(x_{i+1},s_{i+1})$.
This is more accurate than the approach outlined in \citet{13teaetal}, who effectively assume the product $x s(x)$ varies linearly with $x$, which neglects disk center emission and incorrectly weights radiance within each annulus.
A linear approximation of $s(x)$ is preferred over more complex functional forms (e.g., quadratics and cubics) as it avoids over/undershoots between grid points.
Eqn.~(\ref{eqn:approx}) evaluates to:
\begin{eqnarray}
\bar{s} &\approx& \frac{2}{r^2} \sum_{i=0}^{n-2} \left[ \frac{x^2}{2} \left( s_i - \left( \frac{s_{i+1}-s_i}{x_{i+1}-x_i} \right) x_i \right) +\frac{x^3}{3}\left( \frac{s_{i+1}-s_i}{x_{i+1}-x_i} \right) \right]_{x_i}^{x_{i+1}}
\end{eqnarray}
which after grouping terms by $s_i$ and $s_{i+1}$, factorization, and simplification gives:
\begin{eqnarray}
\bar{s} &\approx& \frac{1}{3r^2} \sum_{i=0}^{n-2} s_{i+1}(2x_{i+1}+x_i)(x_{i+1}-x_i) + s_i(x_{i+1}-x_i)(x_{i+1}+2x_i)
\end{eqnarray}
Expanding the sum for annuli $i=0\ldots n-2$:
\begin{eqnarray}
i&=0.   &: \frac{1}{3r^2} \left[ s_{1}(2x_{1}+x_0)(x_{1}-x_0) + s_0(x_{1}-x_0)(x_{1}+2x_0) \right]\\
i&=1.   &: \frac{1}{3r^2} \left[ s_{2}(2x_{2}+x_1)(x_{2}-x_1) + s_1(x_{2}-x_1)(x_{2}+2x_1) \right]\\
i&=2    &: \frac{1}{3r^2} \left[ s_{3}(2x_{3}+x_2)(x_{3}-x_2) + s_2(x_{3}-x_2)(x_{3}+2x_2) \right]\\
 &\ldots& \nonumber \\
i&=n-2 &: \frac{1}{3r^2} \left[ s_{n-1}(2x_{n-1}+x_{n-2})(x_{n-1}-x_{n-2}) + s_{n-2}(x_{n-1}-x_{n-2})(x_{n-1}+2x_{n-2}) \right]
\end{eqnarray}
Grouping terms by $s_i$ gives an expression for $\bar{s}$ as a weighted sum of $s_i$:
\begin{eqnarray}
\label{eqn:wsum}
\bar{s} &\approx& \sum_{i=0}^{n-1} w_i s_i
\end{eqnarray}
where the weights $w_i$ are given by:
\begin{eqnarray}
\label{eqn:w0}
w_0       &=& \frac{1}{3r^2} (x_1-x_0)(x_1+2x_0) ]\\
\label{eqn:wi}
w_i        &=& \frac{1}{3r^2} \left[ (2x_i+x_{i-1})(x_i-x_{i-1}) + (x_{i+1}-x_i)(x_{i+1}+2x_i) \right] \hspace{1cm} \textrm{For } i=1\ldots n-2\\
\label{eqn:wlast}
w_{n-1} &=& \frac{1}{3r^2} (2x_{n-1}+x_{n-2})(x_{n-1}-x_{n-2}) 
\end{eqnarray}
The final disk-averaged spectrum is calculated using Eqn.~(\ref{eqn:wsum}--\ref{eqn:wlast}) from composite spectra $s(x_i)$, $i=0\ldots n-1$.
Note, for a planet of radius $R$, on-disk points ($x<R$) are nadir spectra with emission angles $\theta=\sin^{-1}(x/R)$ and off-disk points ($x\ge R$) are limb spectra with tangent altitudes $x-R$.
Component spectra $s(x_i)$ can be generated using standard radiative transfer codes such as NEMESIS \citep{08irwetal}, archNEMESIS \citep{25aletal}, or PSJ \citep{18viletal}.

\section{Practical considerations}

\subsection{Offset grid $x_i$}

Choice of a sensible $x_i$ grid is critical to the effectiveness of the technique. First, generate a fine (e.g., 5~km) grid of spectra from $x$=0~km to well off the planet limb. The first grid point must be $x_0$=0. Second, identify a sampling scheme that is coarse over smoothly varying offsets and finer where radiance structure occurs. This usually requires finely sampling the limb region, where significant limb brightening can occur. Third chose the maximum offset $r$ to be sufficiently off-limb for the radiance to be negligible compared to measurement noise. The number of grid points $n$ can be chosen by starting from $n$=5 and doubling until the difference between subsequent disk-averaged spectra is small compared to measurement noise.

\subsection{Field-of-view response}

If the instrument field-of-view response varies significantly over the planet, then simply multiply the weights $w_i$ by response function $f(x_i)$ and renormalize weights by $\sum_{i=0}^{n-1} f(x_i)$. Note, the field-of-view must be symmetric about the planet center for this to be valid.

\subsection{Spatially varying emission}

This simplified technique is not suitable if emission has significant spherical asymmetry. For example if a planet's rotation rate ($\Omega$) is such that Doppler shift is significant compared to the spectral resolution $\Delta \lambda$. A valid application requires $\lambda \Omega R/c \ll \Delta \lambda$ where $c$ is speed of light.
Such cases are best handled on a fine 2D grid as in \citet{22teaetal}.

\section{Example application}

\begin{figure*}[ht!]
\plotone{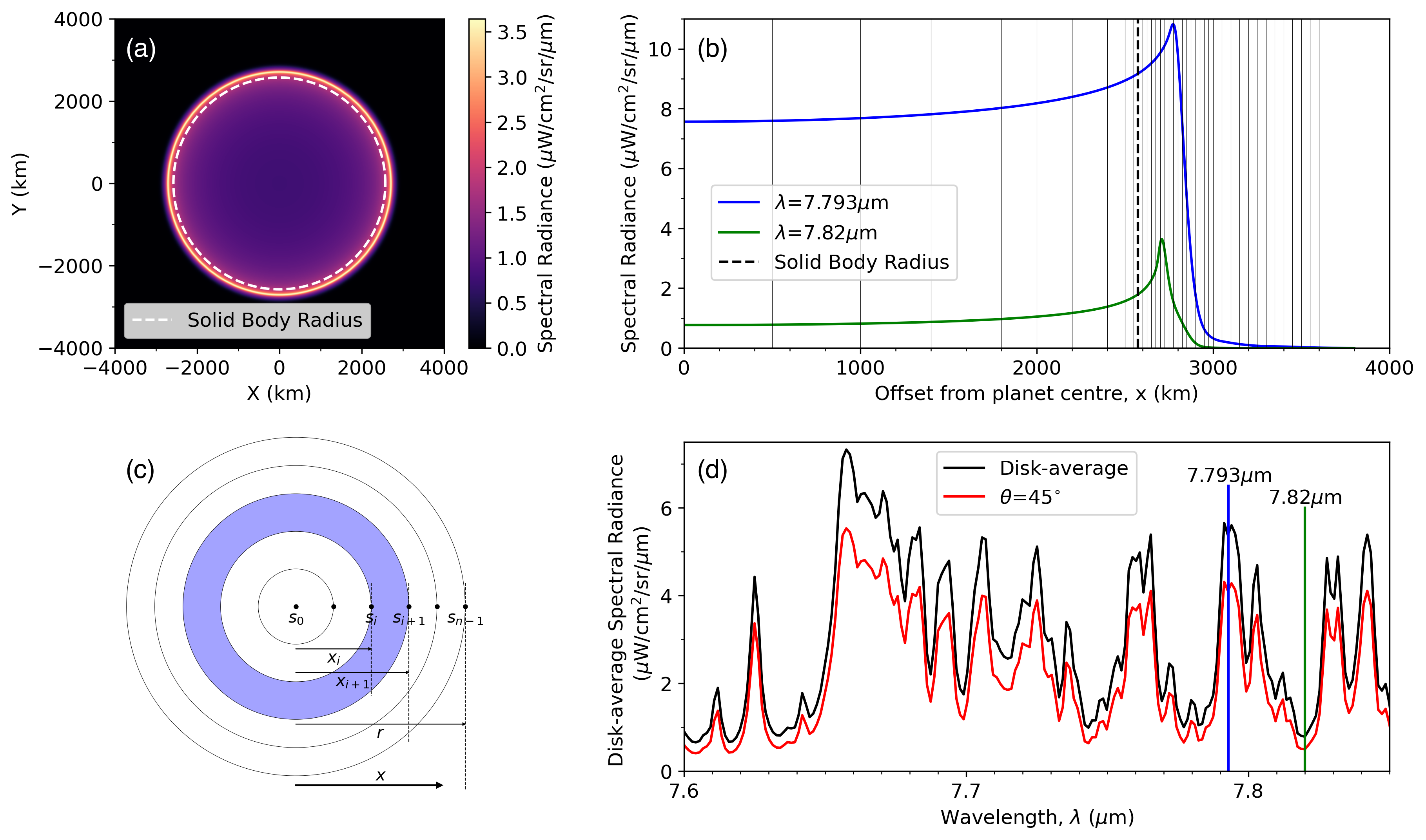}
\caption{Example disk-average spectrum of Titan ($R$=2575~km). (a) Modeled emission at 7.82~$\mu$m. (b) Center-to-limb radiance variation at two example wavelengths. Note significant off-limb emission. (c) Schematic of notation used in this research note. (d) Disk-average spectrum with $n$=40 compared to a modeled 45$^{\circ}$ emission angle spectrum. Vertical lines in (b) indicate the 40-point $x$ grid. The data and code behind this figure are available from \url{https://iopscience.iop.org/article/10.3847/2515-5172/ae8dc7}.}
\label{fig:disk_average}
\end{figure*}

Figure~\ref{fig:disk_average} shows an example application for Titan's 7.7~$\mu$m methane emission band, using synthetic spectra generated using NEMESIS \citep{08irwetal} with standard atmospheric and spectroscopic parameters \citep{19teaetal} and assuming spectral resolution of JWST/MIRI \citep{21labetal}.
Figure~\ref{fig:disk_average} also shows a 45$^{\circ}$ emission angle spectrum for comparison, rescaled by $R^2/r^2$ to account for averaging area differences, which is often used as a simple proxy for a disk-average \citep[e.g.][]{25nixetal}.
Differences between this approximation and the disk-averaging method presented here can be significant.

\begin{acknowledgments}
NAT is funded by UK Science and Technology Facilities Council grant ST/Y000676/1.
\end{acknowledgments}

\software{NEMESIS \citep{08irwetal} ({\tt https://github.com/nemesiscode/radtrancode}))
          }




\begin{thebibliography}{}
\expandafter\ifx\csname natexlab\endcsname\relax\def\natexlab#1{#1}\fi
\providecommand{\url}[1]{\href{#1}{#1}}
\providecommand{\dodoi}[1]{doi:~\href{http://doi.org/#1}{\nolinkurl{#1}}}
\providecommand{\doeprint}[1]{\href{http://ascl.net/#1}{\nolinkurl{http://ascl.net/#1}}}
\providecommand{\doarXiv}[1]{\href{https://arxiv.org/abs/#1}{\nolinkurl{https://arxiv.org/abs/#1}}}

\bibitem[{J. {Alday} {et~al.}(2025){Alday}, {Penn}, {Irwin}, {Mason}, \&
  {Yang}}]{25aletal}
{Alday}, J., {Penn}, J., {Irwin}, P. G.~J., {Mason}, J.~P., \& {Yang}, J. 2025,
  \bibinfo{title}{{archNEMESIS: an open-source Python package for analysis of
  planetary atmospheric spectra},} arXiv e-prints, arXiv:2501.16452,
  \dodoi{10.48550/arXiv.2501.16452}

\bibitem[{T. {Hearty} {et~al.}(2009){Hearty}, {Song}, {Kim}, \&
  {Tinetti}}]{09heaetal}
{Hearty}, T., {Song}, I., {Kim}, S., \& {Tinetti}, G. 2009,
  \bibinfo{title}{{Mid-Infrared Properties of Disk Averaged Observations of
  Earth with AIRS},} Astrophys. J., 693, 1763,
  \dodoi{10.1088/0004-637X/693/2/1763}

\bibitem[{P.~G.~J. Irwin {et~al.}(2008)Irwin, Teanby, de~Kok, Fletcher, Howett,
  Tsang, Wilson, Calcutt, Nixon, \& Parrish}]{08irwetal}
Irwin, P. G.~J., Teanby, N.~A., de~Kok, R., {et~al.} 2008, \bibinfo{title}{The
  {NEMESIS} planetary atmosphere radiative transfer and retrieval tool,} J.
  Quant. Spectro. Rad. Trans., 109, 1136

\bibitem[{A. {Labiano} {et~al.}(2021){Labiano}, {Argyriou},
  {{\'A}lvarez-M{\'a}rquez}, {Glasse}, {Glauser}, {Patapis}, {Law}, {Brandl},
  {Justtanont}, {Lahuis}, {Mart{\'\i}nez-Galarza}, {Mueller}, {Noriega-Crespo},
  {Royer}, {Shaughnessy}, \& {Vandenbussche}}]{21labetal}
{Labiano}, A., {Argyriou}, I., {{\'A}lvarez-M{\'a}rquez}, J., {et~al.} 2021,
  \bibinfo{title}{{Wavelength calibration and resolving power of the JWST MIRI
  Medium Resolution Spectrometer},} Astron. Astrophys., 656, A57,
  \dodoi{10.1051/0004-6361/202140614}

\bibitem[{C.~A. {Nixon} {et~al.}(2025){Nixon}, {B{\'e}zard}, {Cornet}, {Coy},
  {de Pater}, {Es-Sayeh}, {Hammel}, {Lellouch}, {Lombardo},
  {L{\'o}pez-Puertas}, {Lora}, {Rannou}, {Rodriguez}, {Teanby}, {Turtle},
  {Achterberg}, {Alvarez}, {Davies}, {de Kleer}, {Doppmann}, {Fletcher},
  {Hayes}, {Holler}, {Irwin}, {Jordan}, {King}, {Kutsop}, {Marlin}, {Melin},
  {Milam}, {Molter}, {Moore}, {Nyffenegger-P{\'e}r{\'e}}, {O'Donoghue},
  {O'Meara}, {Rafkin}, {Roman}, {Rostopchina}, {Rowe-Gurney}, {Schmidt},
  {Schmidt}, {Sotin}, {Stallard}, {Stansberry}, \& {West}}]{25nixetal}
{Nixon}, C.~A., {B{\'e}zard}, B., {Cornet}, T., {et~al.} 2025,
  \bibinfo{title}{{The atmosphere of Titan in late northern summer from JWST
  and Keck observations},} Nature Astronomy, 9, 969,
  \dodoi{10.1038/s41550-025-02537-3}

\bibitem[{N.~A. {Teanby} {et~al.}(2022){Teanby}, {Irwin}, {Sylvestre}, {Nixon},
  \& {Cordiner}}]{22teaetal}
{Teanby}, N.~A., {Irwin}, P.~G.~J., {Sylvestre}, M., {Nixon}, C.~A., \&
  {Cordiner}, M.~A. 2022, \bibinfo{title}{{Uranus's and Neptune's Stratospheric
  Water Abundance and Vertical Profile from Herschel-HIFI},} Planetary Sci. J.,
  3, 96, \dodoi{10.3847/PSJ/ac650f}

\bibitem[{N.~A. {Teanby} {et~al.}(2019){Teanby}, {Sylvestre}, {Sharkey},
  {Nixon}, {Vinatier}, \& {Irwin}}]{19teaetal}
{Teanby}, N.~A., {Sylvestre}, M., {Sharkey}, J., {et~al.} 2019,
  \bibinfo{title}{{Seasonal Evolution of Titan's Stratosphere During the
  Cassini Mission},} Geophys. Res. Lett., 46, 3079,
  \dodoi{10.1029/2018GL081401}

\bibitem[{N.~A. {Teanby} {et~al.}(2013){Teanby}, {Irwin}, {Nixon}, {Courtin},
  {Swinyard}, {Moreno}, {Lellouch}, {Rengel}, \& {Hartogh}}]{13teaetal}
{Teanby}, N.~A., {Irwin}, P.~G.~J., {Nixon}, C.~A., {et~al.} 2013,
  \bibinfo{title}{{Constraints on Titan's middle atmosphere ammonia abundance
  from Herschel/SPIRE sub-millimetre spectra},} Plan. \& Space Sci., 75, 136,
  \dodoi{10.1016/j.pss.2012.11.008}

\bibitem[{G. {Tinetti} {et~al.}(2005){Tinetti}, {Meadows}, {Crisp}, {Fong },
  {Velusamy}, \& {Snively}}]{05tinetal}
{Tinetti}, G., {Meadows}, V.~S., {Crisp}, D., {et~al.} 2005,
  \bibinfo{title}{{Disk-Averaged Synthetic Spectra of Mars},} Astrobiology, 5,
  461, \dodoi{10.1089/ast.2005.5.461}

\bibitem[{G.~L. {Villanueva} {et~al.}(2018){Villanueva}, {Smith}, {Protopapa},
  {Faggi}, \& {Mandell}}]{18viletal}
{Villanueva}, G.~L., {Smith}, M.~D., {Protopapa}, S., {Faggi}, S., \&
  {Mandell}, A.~M. 2018, \bibinfo{title}{{Planetary Spectrum Generator: An
  accurate online radiative transfer suite for atmospheres, comets, small
  bodies and exoplanets},} J. Quant. Spectro. Rad. Trans., 217, 86,
  \dodoi{10.1016/j.jqsrt.2018.05.023}

\end{thebibliography}
\end{document}